# Urban chaos and replacement dynamics in nature and society

Yanguang Chen

(Department of Geography, College of Urban and Environmental Sciences, Peking University, Beijing 100871, P.R.China. E-mail: chenyg@pku.edu.cn)

**Abstract**: Many growing phenomena in both nature and society can be predicted with sigmoid function. The growth curve of the level of urbanization is a typical S-shaped one, and can be described by using logistic function. The logistic model implies a replacement process, and the logistic substitution suggests non-linear dynamical behaviors such as bifurcation and chaos. Using mathematical transform and numerical computation, we can demonstrate that the 1-dimensional map comes from a 2-dimensional two-group interaction map. By analogy with urbanization, a general theory of replacement dynamics is presented in this paper, and the replacement process can be simulated with the 2-dimansional map. If the rate of replacement is too high, periodic oscillations and chaos will arise, and the system maybe breaks down. The replacement theory can be used to interpret various complex interaction and conversion in physical and human systems. The replacement dynamics provides a new way of looking at Volterra-Lotka's predator-prey interaction, man-land relation, and dynastic changes resulting from peasant uprising, and so on.
**Key words**: bifurcation; chaos; fractal dimension; logistical map; replacement dynamics; Volterra-Lotka's model; urbanization

# 1 Introduction

Logistic function typically reflects a kind of replacement in nature and society, and can be used as one of the substitution models. In early literature, the law of logistic substitution was mainly researched in economics, especially in industrial and technological studies. Fisher and Pry (1971) once successfully used the logistic function to characterize the substitution of old technology by new. Hermann and Montroll (1972) argued that the industrial revolution is a replacement process, that is, the proportion of agricultural worker declines while that of nonagricultural workers increases. Montroll (1978) generalized the notion of substitution to a variety of situations by asserting technological and social evolution to be the consequence of a sequence of substitutions of one technique by another. Treating technological innovations as structural fluctuations in a self-organizing industrial system, Batten (1982) and Karmeshu *et al* (1985) gave a conceptual rationale for Fisher-Pry's law of technological replacement. A landmark of the replacement dynamics is the work of Karmeshu (1988) and his co-workers, who extended the idea of replacement process for the study of replacement of rural population by urban population and revealed the pattern of urbanization in India. Recently, Chen (2009) and Chen and Xu (2010) demonstrated that the replacement dynamics of urbanization may be associated with complex dynamics such as bifurcation and chaos.

Today, it is time to develop a general theory of replacement dynamics by means of the idea from fractals and chaos. The replacement is ubiquitous in both nature and society. Where there is a logistic growth, there is a logistic substitution, and thus replacement behavior will arise. The new points of this work lie in three aspects. First, a general principle of replacement dynamics is proposed. Second, the two-group interaction model is employed to interpret the process of replacement. Third, the periodic oscillations and chaos of replacement dynamics are used to explain the catastrophic occurrences in natural and social evolution. The rest part of this paper is organized as follows. In Section 2, the replacement of urbanization dynamics is advanced, and the 1-dimensional logistic map is linked with the 2-dimensional map of two-population interaction map. In Section 3, the model of replacement dynamics is generalized to different fields such as ecology, geography, and history. In Section 4, a general theory of replacement dynamics is propounded, and the related questions are discussed. Finally, the paper is concluded by summarizing the mains of this work.

## 2 Model: from 1-D map to 2-D map

### 2.1 The 1-D logistic map

The *level of urbanization* is the basic measurement which is used to describe the extent of urbanization in a region. It is defined with the ratio of urban population to total population, that is

$$L(t) = \frac{u(t)}{P(t)} \times 100\% = \frac{u(t)}{r(t)+u(t)} \times 100\%, \tag{1}$$

where $u(t)$ refers to the urban population of time $t$, $r(t)$ to the rural population at the same time, $P(t)$ =$u(t)$+$r(t)$ is the total population of the region, and $L(t)$, the level of urbanization. The measurement equivalent to this is the *urban-rural ratio* (URR), which is defined by $o(t)=u(t)/r(t)$ (Chen and Xu, 2010; United Nations, 2004). Don't look down on these simple measurements. Just due to them, the relation between the 1-dimensional logistic map and a 2-dimensional two-population interaction map is brought to light (Chen, 2009).

In urban studies, one of problems concerning us is how the level of urbanization changes over time. In fact, if some kind of measurement of a system has clear upper limit and lower limit, the growing course of the measurement always takes on the S-shaped curve. The curve can be formulated as a *sigmoid function*. The sigmoid function is also called *squashing function*. Pressed by the upper limit and withstood by the lower limit, a line will be twisted into S shape. The family of sigmoid functions includes various functions such as the ordinary arc-tangent, the hyperbolic tangent, and the generalized logistic function. Among all these sigmoid function, the simplest and the best-known one is the logistic function. Sometimes, the logistic function is the synonym of the sigmoid function (Mitchell, 1997).

As we know, the level of urbanization comes between 0 (or 0%) to 1 (or 100%), thus it can often be fitted to the logistic function since it has clear upper and lower limits (Chen, 2009; Chen and Xu, 2010; Karmeshu, 1988; Rao *et al*, 1989). For many years, the United Nations experts of urbanization employed the logistic model to predict the level of urbanization of different countries (United Nation, 2004). The logistic model can be expressed as

$$L(t) = \frac{1}{1+(1/L_0 - 1)e^{-kt}}, \tag{2}$$

where $L_0$ is the initial value of urbanization level $L(t)$, i.e., the level of urbanization of time $t$=0, $k$ is the intrinsic/ original growth rate. The derivative of equation (2) is

$$\frac{dL(t)}{dt} = kL(t)[1-L(t)]. \tag{3}$$

Obviously, this is a discrete-time demographic model analogous to the logistic equation first created by mathematician Pierre François Verhulst (Banks, 1994). Suppose that the step length of data sampling is Δ*t*=1. Discretizing equation (3) yields a 1-dimensional map in the form

$$L_{t+1} = (1+k)L_t - kL_t^2 . \qquad (4)$$

where $L_t=u_t/(u_t+r_t)$ is the discrete expression of $L(t)$, and here $r_t$ and $u_t$ are the discrete expressions of $r(t)$ and $u(t)$, respectively.

The logistic map is in fact a polynomial mapping of degree 2 based on recurrence relation. The map was popularized in a seminal paper of May (1976). It is often cited as an archetypal example of how complex behavior such as bifurcation and chaos can emerge from very simple non-linear dynamical equations (Figure 1). Let $x_t=kL_t/(1+k)$, then equation (4) can be normalized and we have $x_{t+1}=(1+k)x_t(1-x_t)$. I will demonstrate that the logistic process is actually a replacement process, which can be extended to the general principle of replacement dynamics.

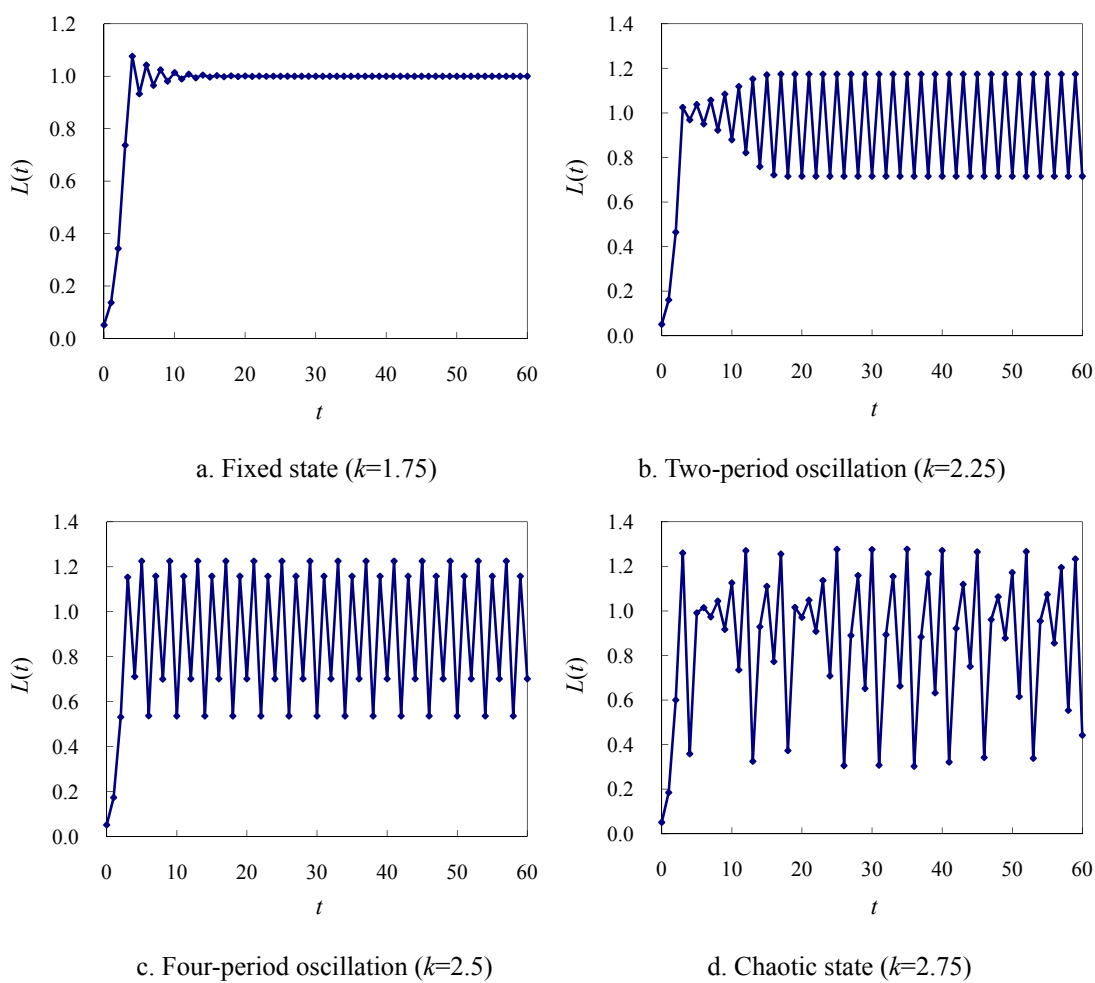

a. Fixed state (*k*=1.75)   b. Two-period oscillation (*k*=2.25)

c. Four-period oscillation (*k*=2.5)   d. Chaotic state (*k*=2.75)

**Figure 1 The change of the level of urbanization based on the 1-dimensional map: from fixed state to chaotic state** (the pattern is the same as those from May (1976))

## 2.2 The 2-D two-population interaction map

Because of equation (1), the 1-dimensional map can be converted into a 2-dimensional map of urban-rural population interaction. The level of urbanization is defined by urban population and rural population, thus the percent of urban population must be associated with urban and rural population growth. By analogy with the Volterra-Lotka model in ecology (Dendrinos and Mullally, 1985), the urban-rural interaction model can be built as follows (Chen, 2009)

$$\begin{cases} \dfrac{dr(t)}{dt} = ar(t) - b\dfrac{r(t)u(t)}{r(t)+u(t)} \\ \dfrac{du(t)}{dt} = cu(t) + d\dfrac{r(t)u(t)}{r(t)+u(t)} \end{cases}. \tag{5}$$

This means that the growth rate of rural population, d$r(t)$/d$t$, is proportional to the size of rural population, $r(t)$, and the coupling between rural and urban population, but not directly related to urban population size; the growth rate of urban population, d$u(t)$/d$t$, is proportional to the size of urban population, $u(t)$, and the coupling between rural and urban population, but not directly related to rural population size. It seems to be difficult to understand these, but the census dataset of the United States (US) of America supports this model (Chen and Xu, 2010). What is more, the two-population interaction can be generalized to two-group interaction.

If a study region is a close system, the decrease of rural population will be equal to the increase of urban population resulting from urban-rural interaction, and thus we have $b=d$. In this instance, from equations (1) and (5) follows

$$\dfrac{dL(t)}{dt} = (b-a+c)L(t)[1-L(t)]. \tag{6}$$

Let $k=b-a+c$, equation (6) is identical to equation (3). This suggests that the urbanization level growth is associated with the process of urban and rural interaction. Discretizing equation (5) yields a 2-dimenaional map such as

$$\begin{cases} r_{t+1} = (1+a)r_t - b\dfrac{r_t u_t}{r_t + u_t} \\ u_{t+1} = (1+c)u_t + d\dfrac{r_t u_t}{r_t + u_t} \end{cases}. \tag{7}$$

For simplicity, the notation of parameters is not changed in spite of the error coming from the continuous-discrete conversion. By means of the least squares calculation of the US census data

from 1790 to 1960, we have

$$\begin{cases} r_{t+1} = 1.2584 r_t - 0.3615 \dfrac{r_t u_t}{r_t + u_t} \\ u_{t+1} = u_t + 0.5044 \dfrac{r_t u_t}{r_t + u_t} \end{cases}.$$

I don't adopt the data after 1960 as the US city definition varied from 1970.

US is not a close country of population, so it is not strange that $b\approx0.3615$ is not equal to $d\approx0.5044$. What is strange is that $c=0$, this suggests that the growth rate of urban population is only proportional to the coupling between rural and urban population, but not directly related to rural and urban population sizes. However, if we think it deeply, it seems to be true. Despite cities of long standing, the history of urbanization compared with the history of human being is not long. There was no real urbanization before industrialization (Knox, 2005). If the growth rate of urban population was proportional to its urban population size, urbanization should arise long ago. As a matter of fact, urbanization began due to industrialization, development of transportation and so on. Numerical simulation shows that, if $c$ is significantly greater than 0, urban population and total population in a region will grow ceaselessly. However, this is not the case in the real world. Therefore, the $c$ value is either zero or very small.

Table 1 The model parameters of urbanization and corresponding dynamical behaviors ($a=0.25$, $c=0$)

| Figure | 1-D map ($k=b-a+c$) | 2-D map ($b=d$) | System behavior |
|---|---|---|---|
| Figure 1(a), Figure 2(a) | 1.75 | 2.00 | Fixed state |
| Figure 1(b), Figure 2(b) | 2.25 | 2.50 | Two-period oscillation |
| Figure 1(c), Figure 2(c) | 2.50 | 2.75 | Four-period oscillation |
| Figure 1(d), Figure 2(d) | 2.75 | 3.00 | Chaotic state |

The nonlinear dynamical behaviors displayed in Figure 1 can also be generated by the 2-dimensional map, equation (7). According to the US model of urbanization, let $a=0.025$, $c=0$ (Table 1). For simplicity and without loss of generality, let $b=d$. the US census started in 1790, when the urbanization is 201 655, and the rural population is 3 727 559. We may take the initial values such as $r_0=3.727\,559$ million, and $u_0=0.201\,655$ million. Thus, the $r_t$ and $u_t$ values can be given by the recurrence relations, equation (7), and the level of urbanization, $L_t$, can be given by equation (1), or $L_t=u_t/(u_t+r_t)$. Changing the *b* and *d* values, we have various simple and complex dynamical patterns such as sigmoid growth, periodic oscillations, and chaos (Figure 2). In fact, if we remove various constraints imposed on the parameter *a*, *b*, *c*, and *d*, the dynamical behaviors of the 2-dimenaional map are richer and more colorful than those of the 1-dimensiaonl map.

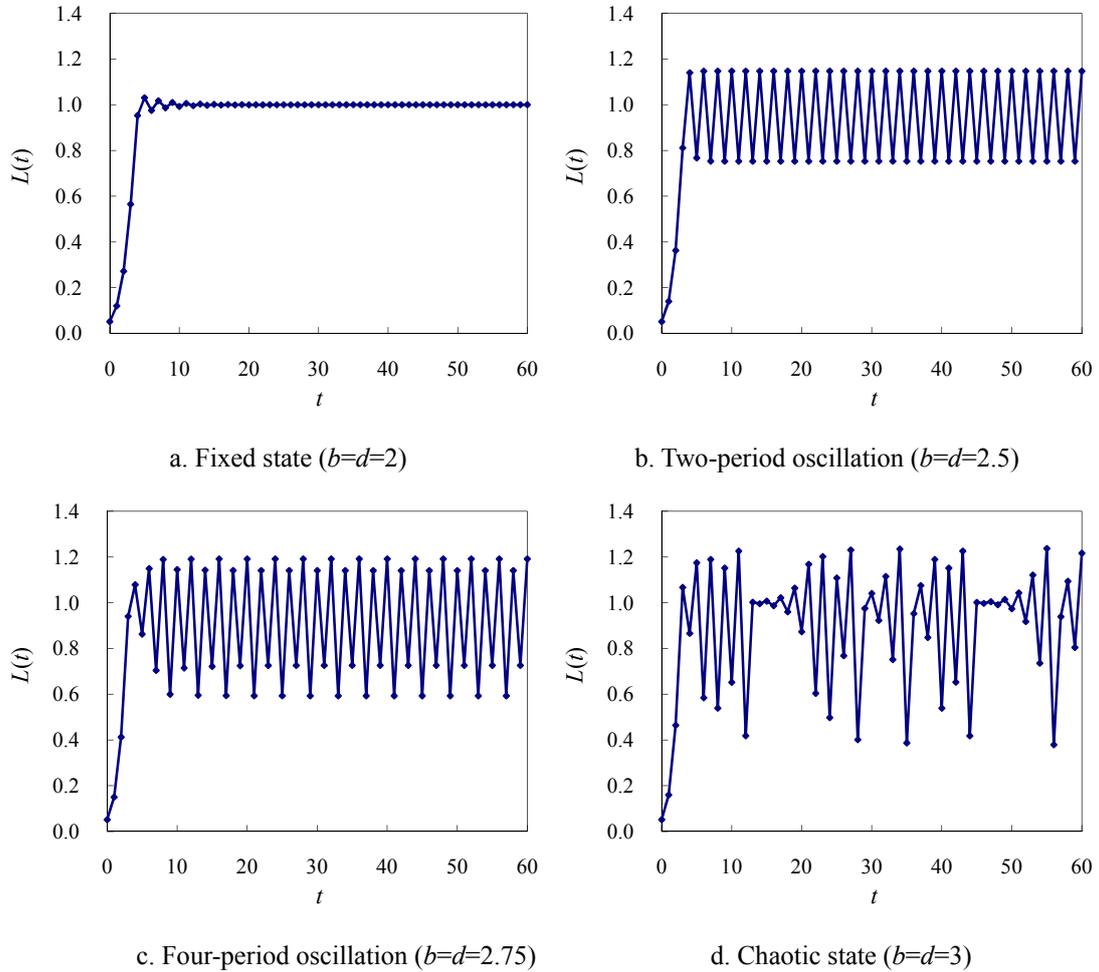

a. Fixed state (*b*=*d*=2)    b. Two-period oscillation (*b*=*d*=2.5)

c. Four-period oscillation (*b*=*d*=2.75)    d. Chaotic state (*b*=*d*=3)

**Figure 2 The change of the level of urbanization based on the 2-dimensional map: from fixed state to chaotic state (Chen, 2009)**

### 3.3 Replacement dynamics

Urbanization is in fact a process of population replacement—the urban population substitute for the rural population (Karmeshu, 1988). Generally speaking, the replacement equation is an exponential function, can the replacement process can be expressed as a logit transform. As indicated above, URR is defined by

$$O(t) = \frac{u(t)}{r(t)}. \tag{8}$$

The level of urbanization can be expressed as

$$L(t) = \frac{u(t)}{u(t)+r(t)} = \frac{1}{1+1/O(t)}. \quad (9)$$

In light of equation (2), URR proved to be an exponential function of time, namely

$$O(t) = \frac{L(t)}{L(t)-1} = (\frac{L_0}{L_0-1})e^{kt} = O_0 e^{kt}. \quad (10)$$

where $O_0 = L_0/(L_0-1)$ is the initial value of URR. Thus we have logit transform such as

$$\ln O(t) = \ln \frac{L(t)}{1-L(t)} = \ln O_0 + kt = \ln \frac{L_0}{1-L_0} + kt, \quad (11)$$

which is equivalent to

$$\ln \frac{u(t)}{r(t)} = \ln \frac{u_0}{r_0} + kt, \quad (12)$$

where the initial values $r_0 = r(0)$, $u_0 = u(0)$. This is just the urban-rural population substitute equation (Karmeshu, 1988). Both the urban-rural population datasets of American census and Indian census can be fitted to the replacement model, equation (10) or equation (12) (Figure 3). Moreover, the population replacement can be extended to the general replacement dynamics.

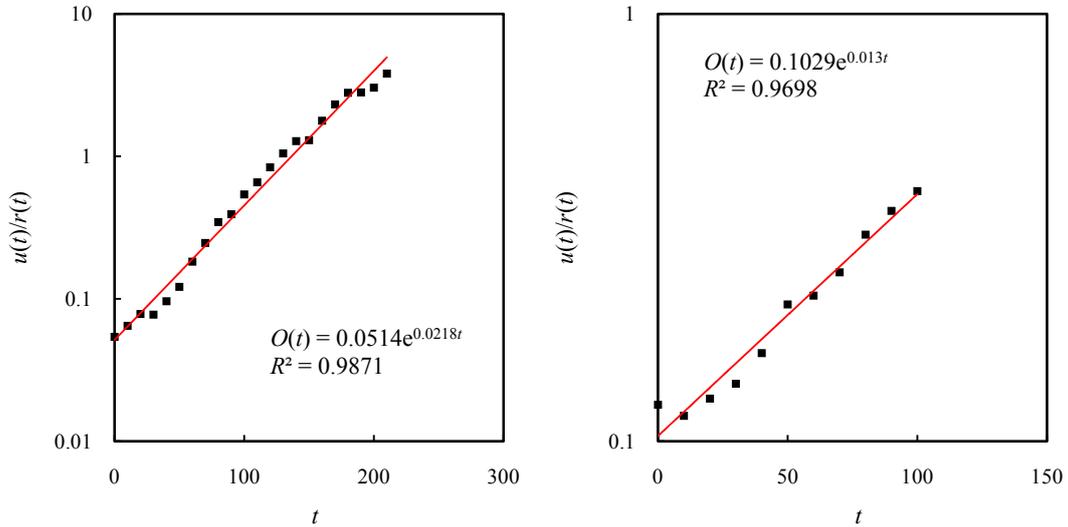

a. America, 1790-2000      b. India, 1901-2001

**Figure 3 Two patterns of the urban replacement process of America and India**

(**Note**: The data sources are censuses of the United States of America and India, respectively. See: (1) http://www.census.gov/population; (2) http://censusindia.gov.in/Census_Data_2001/ .)

# 3 Generalization: the replacement dynamics in nature and society

## 3.1 Example 1: ecological replacement

The urban replacement came from technological replacement, can be generalized to many different fields. Let's see several examples. In ecology, the predator-prey interaction model is a well-known nonlinear dynamical model (Dendrinos and Mullally, 1985). The model can be revised by using the concept from replacement dynamics. Suppose there a close region such as an island with two types of animals. One is predator (flesh-eater, carnivore, carnivorous animals), and the other, the prey (vegetarian, herbivore, herbivorous animals). By analogy with the energy unit "standard coal", we can make a comparable unit for animals, namely, *standard head*. If the predators and preys are $x$ and $y$ standard heads, respectively, then we can define a measurement, the ratio of the biomass of predator to that of the entire animals $p(t)$, in the form

$$p(t) = \frac{x(t)}{x(t) + y(t)} \times 100\% . \tag{13}$$

where $x$ represents the biomass of predator, $y$, the biomass of prey in the region. Accordingly, the predator-prey ratio (PPR) can be given by $o(t)=x(t)/y(t)$.

Clearly, there is upper limit and lower limit for $p(t)$. Because of the *squashing effect*, the ratio of the biomass of predator to that of the entire animals should follow the law of logistic growth. Thus, Volterra-Lotka's predator-prey interaction model can be revised in the form of equation (5), and May (1976)'s logistic equation can be associated with the revised predator-prey model, which differ in expression from the classical Volterra-Lotka's model.

## 3.2 Example 2: geographical replacement

In geography, the very basic and important topic is the relation between man and land or natural environment. Because of the man-land interaction, the primary productivity or even secondary productivity are gradually consumed by human being. So far, man has used up more than 40% of the primary productivity. In other words, human being has transformed the first nature into the second nature and the third nature (Swaffield, 2002). Defined a new measurement, the ratio of the primary productivity to the total productivity in a region $p'(t)$, in the form

$$p'(t) = \frac{x'(t)}{x'(t) + y'(t)} \times 100\% . \tag{14}$$

where *x'* represents the primary productivity, and *y'* is the other productivity. Accordingly, the ratio of the primary productivity to the other productivity (POR) can be defined by $o(t)' = x'(t)/y'(t)$.

The process of human being marauding nature is also a logistic process, and thus it is also a nonlinear dynamics of replacement. In a sense, the man-land relation is unstable, and periodic oscillations and chaos may arise since humankind depletes nature so rapidly.

### 3.3 Example 3: historical replacement

In the theory of class struggle, Marxist analysis of society identifies two main social groups: *labor* (the proletariat or workers) and *capital* (the bourgeoisie or capitalists). The former includes anyone who earns their livelihood by selling their labor power and being paid a wag/salary for their labor time, and the latter include anyone who obtains their income not from labor as much as from the surplus value that they appropriate from the workers who create wealth. Whether or not you believe in Marxism, you will agree that the social system falls into two groups—the haves (wealthy, men of wealth, the rich) and the have-nots (needy, poor people, the poor). Let *x"* represent the number of have-nots, and *y"* represent the number of the haves. Then we can define a ratio of the have-nots to the total population such as

$$p''(t) = \frac{x''(t)}{x''(t) + y''(t)} \times 100\% . \tag{15}$$

Thus the poor-rich ratio (PRR) can be expressed as $o''(t) = x''(t)/y''(t)$. This process of replacement is just so-called pauperization of a country.

In ancient China, for example, dynastic changes were always processes of land annexation and concentration of landholding right. The rich annexed the land of the poor, first little by little, and then more and more, so that about 20 percent of people held 80 percent of land or so, and about 80 percent people had little land. Land annexation can be regarded as a logistic process. It often led to social pauperization, and then the nation lost its balance. Peasant uprising burst out, social turbulence began. Hundreds responded to a single call, and as one (uprising) fell, another rose. This seems to be the course from oscillations to chaos. The final result is that new dynasty replaced old dynasty, or new regime replaced old regime.

# 4 Questions and discussion

The logistic substitute model can be employed to describe the process of replacement of one activity by another (Karmeshu, 1988). The studies used to be confined in the fields of industrial development and urbanization. However, you can find the replacement processes here and there in both nature and society. If some phenomena fill into two groups, which can be expressed by the dummy variable 0 and 1, and one group tries to take the place of the other group, there will be a replacement dynamics (Table 2). So, the replacement dynamics is actually a 0-1 discrete spatial dynamics. The living space of one group is squashed and occupied by another group. The replacement process can be divided into two types: *virtuous replacement* (sound substitution) and *vicious replacement* (unsound substitution). For examples, new technology displace old technology, this is a kind of virtuous replacement. However, the land annexation in human history (the have-nots displace the haves) is a kind of vicious replacement. The virtuous replacement lead to a virtuous cycle, and a system develops soundly, whereas the vicious replacement result in a vicious cycle, and a system tends towards breakdown.

Table 2 The 0-1 classification of living space and the elements of replacement dynamics

| Item | Type 0 | Type 1 |
|---|---|---|
| **Industry 1** | Old technology | New technology |
| **Industry 2** | Traditional technique | Modern technique |
| **Urbanization** | Rural population | Urban population |
| **City** | Unfilled space | Filled space |
| **Ecology 1** | Prey | Predator |
| **Ecology 2** | Primary productivity | The secondary and tertiary productivity |
| **Geography 1** | The first nature | The second, third, and fourth nature |
| **Geography 2** | Natural space | Human space |
| **Economics 1** | Primary industry | The secondary and tertiary industry |
| **Economics 2** | Agricultural workers | Non-agricultural workers |
| **Sociology** | Old ideas | New ideas |
| **History** | Haves (the rich) | Have-nots (the poor) |

It is necessary to clarify the dynamical mechanism of the discrete replacement. Let's take the urban replacement as an example to illustrate the underlying rationale of replacement process.

Two models based on two postulates can be employed to explain the replacement dynamics of urbanization. One the nonlinear dynamics model based on the postulate of two-population coupling, as indicated by equation (5), the other is a linear dynamics model based on the postulate of allometric growth such as (Bertalanffy, 1968; Chen and Jiang, 2009)

$$\begin{cases} \dfrac{dr(t)}{dt} = \alpha r(t) \\ \dfrac{du(t)}{dt} = \beta u(t) \end{cases}, \quad (16)$$

where $\alpha$ and $\beta$ are growth coefficients ($\alpha>0$, $\beta>0$). The solutions to linear differential equations are

$$\begin{cases} r(t) = r_0 e^{\alpha t} \\ u(t) = u_0 e^{\beta t} \end{cases}. \quad (17)$$

From equation (17) follows

$$O(t) = \frac{u(t)}{r(t)} = \frac{u_0}{r_0} e^{(\beta-\alpha)t} = \frac{u_0}{r_0} e^{kt} = O_0 e^{kt}, \quad (18)$$

which is equivalent to equation (12). This suggests that the replacement model of urbanization can be derived from the linear dynamics equations. Apparently we have

$$k = \beta - \alpha. \quad (19)$$

Because $k$ represents the difference between the relative growth ratio of rural population, $\alpha$, and that of urban population, $\beta$, the United Nations (2004) experts called it the *urban-rural growth difference* (URGD). On the other hand, from equation (17) follows an allometric scaling relation between urban and rural population in the form

$$u(t) = (\frac{u_0}{r_0^{\beta/\alpha}}) r(t)^{\beta/\alpha} = \eta r(t)^b, \quad (20)$$

where $\eta = u_0/r_0^{\beta/\alpha}$ is proportionality coefficient, and

$$b = \frac{\beta}{\alpha} = \frac{D_u}{D_r}, \quad (21)$$

is the allometric scaling exponent. Equation (20) was verified by Naroll and Bertalanffy (1956). In equation (21), $D_u$ refers to the fractal dimension of urban population, and $D_r$, the fractal dimension of rural population. This suggests that, if the replacement dynamics is based on the linear differential equations, it is actually based on the allometric scaling law associated with fractal geometry. The law of allometric growth is very significant in urban studies (see e.g. Batty and Longley, 1994; Bettencourt *et al*, 2007; Chen, 2010; West et al, 2002). However, in the urban replacement process, it seems to be the non-linear interaction rather than the allometric scaling that dominates the dynamical evolution.

Discretizing equation (16) yields a 2-dimensional map such as

$$\begin{cases} r_{t+1} = (1+\alpha)r_t \\ u_{t+1} = (1+\beta)u_t \end{cases}. \tag{22}$$

However, if we employ equation (22) to simulate the rural population, urban population, total population, and the level of urbanization, the results do not tally with the actual situation—all the population and the urbanization level go up unboundedly. In contrast, if we use the 2-dimensional map defined by equation (7) to simulate the level of urbanization, the rural population, urban population, and total population, the results conform to reality—the capacity values are limited (Figure 4, Figure 5). This seems to imply that it is the nonlinear dynamics rather than the linear dynamics that can be employed to interpret the replacement process. Despite this, the logistic equation can be used as a bridge of understanding between the simple allometric scaling law and the complex nonlinear interaction model (Table 3).

Table 3 Mathematical transform relation between the simple allometric scaling laws and complex non-linear dynamics

| Models | Complex models | Transformation | Simple models |
|---|---|---|---|
| Equations | $\begin{cases} \dfrac{dr(t)}{dt} = ar(t) - b\dfrac{r(t)u(t)}{r(t)+u(t)} \\ \dfrac{du(t)}{dt} = cu(t) + d\dfrac{r(t)u(t)}{r(t)+u(t)} \end{cases}$ | $\dfrac{dL(t)}{dt} = kL(t)[1-L(t)]$ | $\begin{cases} \dfrac{dr(t)}{dt} = \alpha r(t) \\ \dfrac{du(t)}{dt} = \beta u(t) \end{cases}$ |
| Mathematical Solution | No | $L(t) = \dfrac{1}{1+(1/L_0 - 1)e^{-kt}}$ | $\begin{cases} r(t) = r_0 e^{\alpha t} \\ u(t) = u_0 e^{\beta t} \end{cases}$ |
| Numerical Solution | Bifurcation and chaos | No | Simple curve |
| Theory | Bifurcation and chaos | Scaling laws | Fractals |

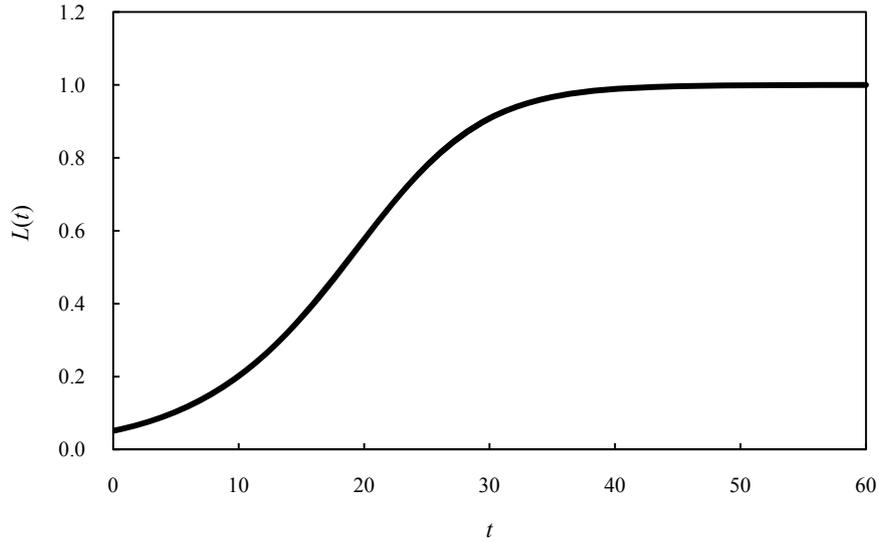

**Figure 4 The logistic curve of urbanization level based on the 2-dimensional map**

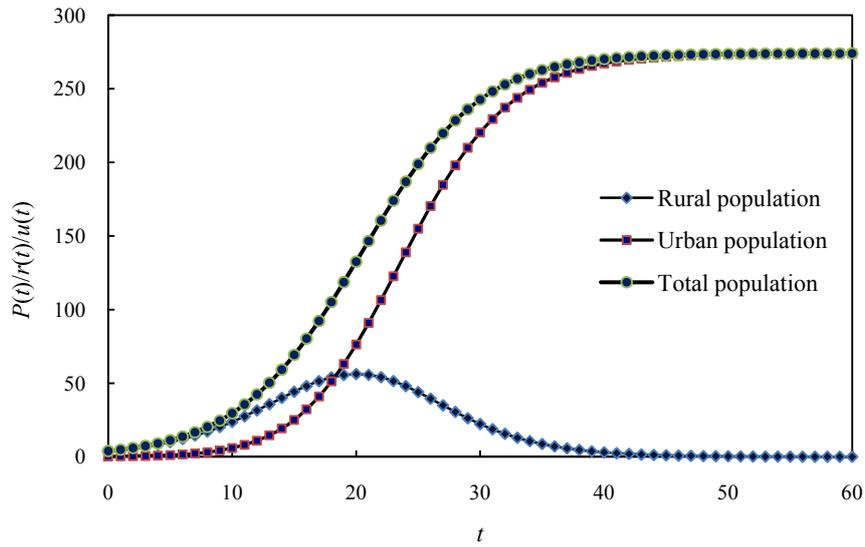

**Figure 5 The growth curves of the urban population, rural population, and total population based on the 2-dimensional map**

The replacement complex depends on the rate of logistic substitution. If the parameter values *a*, *b*, *c*, and *d* in the 2-dimensional map, equation (7), have proper scales, the numerical simulation results are normal, and the rural population, urban population, total population, and the level of urbanization finally converge (Figure 4, Figure 5). However, if the parameter values go beyond certain limits, the periodic oscillation and chaos will emerge. Figure 2 displays the change patterns of the level of urbanization. The pivotal parameters are the ones on the cross terms, that is, the coupling parameters *b* and *d*. The values of *b* and *d* indicate the rate of replacement. The higher the *b* and *d* are, the rapider the replacement is. If and only if *b* and *d* exceed the critical values, periodic oscillations and chaos will arise. A conclusion can be drawn that the overspeed ("quick-tempered") replacement gives rise to unstable evolution.

# 5 Conclusion remarks

The logistic replacement is one of the ubiquitous general empirical observations across the individual sciences, which cannot be understood in the set of references developed within the certain scientific domain. We can find the replacement process everywhere in nature and society. The theory of replacement dynamics should be developed from the interdisciplinary perspective. It deals with the replacement of one activity by another. One typical logistic substitution is the replacement of old technology by new, another typical logistic substitution is the replacement of rural population by urban population. This paper is mainly based on urban-rural replacement, but it provides a series of examples on replacement process in different fields such as ecology, geography, and history.

The replacement process is associated with the nonlinear dynamics described by two-group interaction model. The discrete expression of the nonlinear differential equation is the 2-dimensional maps. The map can generate various simple and complex behaviors including S-shaped growth, periodic oscillations, and chaos. If the rate of replacement is lower, the growth curve is a sigmoid curve. However, if the replacement rate is too high, periodic oscillations or even chaos will arise. This suggests, no matter what kind of replacement it is--virtuous substitution or vicious substitution, the rate of replacement should be befittingly controlled. Otherwise, catastrophic events maybe take place, and the system will likely fall apart. Clearly, the studies on the replacement dynamics is revealing for us to understand the evolution in nature and society.

**Acknowledgements**

This research was sponsored by the National Natural Science Foundation of China (Grant No. 40771061). The support is gratefully acknowledged.